# Interfacial Engineering Enabled High-Resolution Stretchable Metal-Level Conductive Lines and Transparent Conductor

Ke Chen[1]†, Eric Tianjiao Zhao[1]†, Weichen Wang[1,2]†, Yepin Zhao[1]†, Kostas Parkatzidis[1], Lukas Michalek[1], Yuran Shi[1], Elizabeth Zhang[2], Tianyang Chen[1], Ruiheng Wu[1], Hao Lyu[1], Yating Yao[3], Yujia Yuan[4], Tianhao Chen[1], Ruby Rong Zhou[2], Shiyuan Wei[1], Junyi Zhao[1], Jeffrey B.-H. Tok[1], Zhenan Bao[1]*

[1]Department of Chemical Engineering, Stanford University, Stanford, CA 94305, USA.
[2]Department of Materials Science and Engineering, Stanford University; Stanford, CA 94305, USA.
[3]Department of Chemistry, Stanford University; Stanford, CA 94305, USA.
[4]Department of Electrical Engineering, Stanford University; Stanford, CA 94305, USA.

*Correspondence should be addressed to zbao@stanford.edu (Z.B.)

†These authors contributed equally to this work

**Abstract:**
As stretchable electronics advance toward higher integration density and finer feature sizes, stretchable conductors, which serve as the architectural backbone of these electronics, must be scaled down accordingly. However, achieving both high stretchability and high electrical conductivity at high resolution remains a challenge using existing conductive materials. Here, through material design and interfacial engineering, we develop a thiol-functionalized conducting polymer/gold hybrid stack that can be patterned down to 4 μm linewidths using standard photolithography, while maintaining high stretchability, metal-like conductivity (>40,000 S/cm), and environmental stability. Leveraging this capability, we demonstrate grid-based stretchable transparent electrodes that surpass the figure-of-merit of indium tin oxide, as well as a 1000-pixel-per-inch image interconnected with stretchable lines. This work helps to broaden the scope of next-generation functional soft electronics, including e-skins and bioelectronics.

## Main Text

Stretchable conductors are inherent in stretchable electronics. They are required to transmit electrical signals among active components and define the overall design of an electronic system (*1*–*3*). With the continued miniaturization of stretchable electronic systems, including high-density neural probes (*4*, *5*), stretchable high-resolution LED displays (*6*, *7*), microprocessors (*8*, *9*), and sensor chips (*10*), conductors must hence be patterned into high-resolution interconnects to accommodate the increasing complexity of circuitry within a compact footprint. This requires the stretchable conductors to exhibit high electrical conductivity, be compatible with high-resolution patterning techniques, and maintain high stretchability even at narrow line widths (**Fig. 1A**). Additionally, they should exhibit environmental stability to ensure reliable performance.

Various coating and blending strategies have been reported to confer stretchability to rigid metallic conductors by nanostructuring with polymer matrices or strain-engineered structures, e.g., wavy or serpentine geometries (*11*, *12*). However, the coating/blending approach is unable to simultaneously achieve high conductivity and high-resolution patterning, while strain engineering has a trade-off between stretchability and device density. Another approach relies on engineering metal cracks, in which cracks form under strain, but the percolation pathway can nonetheless be maintained (*13*, *14)*. However, this strategy becomes increasingly challenging as feature sizes shrink, because crack dimensions must remain much smaller than the interconnect width to prevent complete electric current disruption. As a result, crack-engineered metals lose reliability at reduced dimensions and have not demonstrated robust stretchability below 25 μm linewidths (*15*). In other approaches, liquid metals such as EGaIn were observed to exhibit superior stretchability and conductivity (*16*). However, they are susceptible to oxidation and chemical reactions in humid and biological environments, leading to unstable electrical performance (*17*). Furthermore, the fluid nature of liquid metals complicates long-term packaging and limits compatibility with dry-state lithography for precise multilayer alignment and integration (*18*, *19*). Besides metallic conductors, intrinsically stretchable conductors, such as Poly(3,4-ethylenedioxythiophene): poly(styrene sulfonate) (PEDOT:PSS), have been explored for their softness and stretchability at small line width (*20*, *21*). Unfortunately, their electrical conductivity remains substantially lower than that of metallic conductors. Thus, we aim to develop a material platform that simultaneously combines metal-level conductivity, high stretchability, high patterning resolution, and environmental stability.

Here, we report a stretchable, highly conductive hybrid PEDOT/Au bilayer material platform that can be stretched to ~170% (**Fig. 1B**) and achieve metal-like level conductivity >40,000 S/cm. It can be patterned down to 4 μm using a standard photolithography process while maintaining electrical performance under 50 % strain, and larger features (10–110 μm) remain electrically functional up to 75%–150% strain. Briefly, this feat was enabled through rational engineering of the PEDOT formulation and controlling the PEDOT/Au interface. We first developed a thiol-rich PEDOT, which can be stretched >150% and still maintain conductivity (**Fig. 1B (ii)**). Next, at the PEDOT/Au interface, the evaporated Au partially diffuses into PEDOT to form an interpenetrating network on the surface and robust bonding between PEDOT through thiol groups (**Fig. 1B (iii)**). The coexistence of physical interlocking and chemical bonding significantly enhances PEDOT/Au interfacial adhesion, enabling the Au layer (~20–30 nm) to develop only submicron-scale cracks even under 170% strain. Since conducting PEDOT ensures continuous charge transport, these features enable us to achieve stable, high electrical conduction at micron-sized line width under strain.

A unique application of this hybrid material is for transparent electronics. We show these high-resolution PEDOT/Au lines can be applied as grid-based transparent electrodes with low sheet resistance (< 20 Ω/sq) and high transmittance (>90%). This is on par with the figure-of-merit in both indium tin oxide and silver nanowire (AgNW) transparent electrodes while having >100% strain stretchability (**Fig. 1C**). We further demonstrate a transparent stretchable micro-LED display (**Fig. 1D**) and 1000-PPI (pixel per inch) patterns with our fabricated interconnects, which are inaccessible with existing stretchable conductors. Unlike EGaIn interconnects, which degrade rapidly in humid air or water, our hybrid PEDOT/Au interconnect shows negligible changes in conductivity and impedance in PBS solutions, humid environment, and in air for >15 months. Taken together, our developed material represents the first to meet all the stringent needs for high-resolution, stable, stretchable, and highly conductive interconnect, while being readily prepared via modern microfabrication lithographic process. Our developed interfacial-engineering approach will facilitate emerging soft electronic systems that demand mechanical deformability, high electrical performance, and stability at small footprints (**Fig. 1E-1F**).

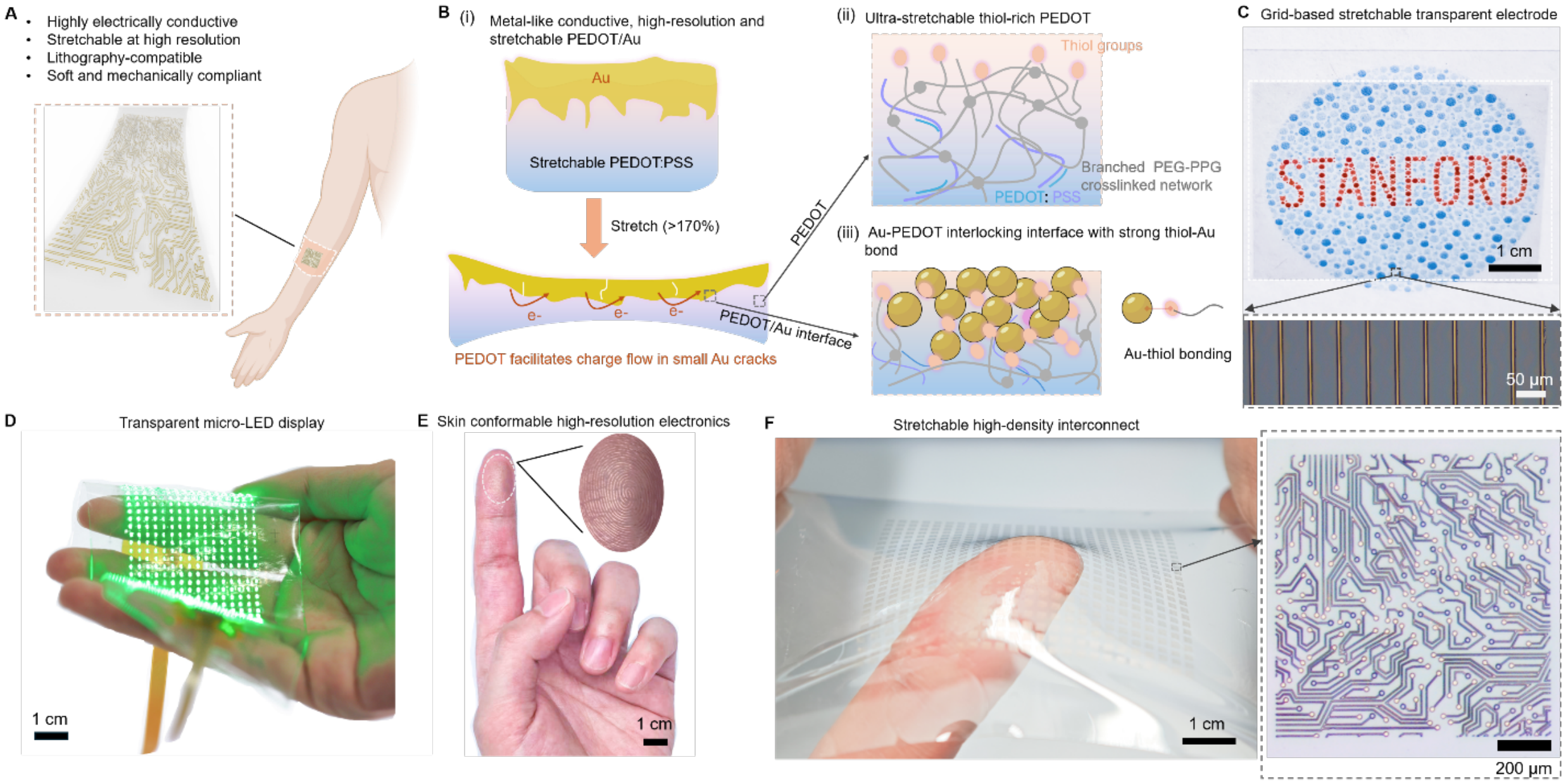

**Figure 1. Stretchable, conductive, and high-resolution thiol-PEDOT/Au interconnects for transparent electronics.** (A) High-resolution interconnects are essential for high-density stretchable electronics. (B) (i) Schematic illustration of high-resolution stretchable PEDOT/Au interconnects. Upon stretching, the conductive thiol-PEDOT layer helps to maintain good charge transport across the micro-cracked Au film. The magnified view shows that: (ii) stretchable thiol-PEDOT consists of PEDOT:PSS crosslinked with a branched PEG–PPG network with abundant thiol functional groups near the surface; (iii) At the thiol-PEDOT/Au interface, Au partially diffuses into PEDOT and forms a strong bonding with the PEDOT matrix through thiol-functional groups. This facilitates robust adhesion between thiol-PEDOT and Au. These features collectively allow the interconnects to maintain high conductivity and stretchability even at micro-sized widths. (C) Photograph of a transparent conductor based on 5 μm PEDOT/Au grids. (D) Photograph of a large-area transparent micro-LED display. (E) Photograph of a high-resolution, skin-conformable PEDOT/Au interconnect with a fingertip pattern. (F) Photograph of a high-density PEDOT/Au interconnect array for integrated circuits. The right image is a magnified optical micrograph of one block in the photo. The interconnect width is 4 μm.

## Stretchable thiol-functionalized PEDOT (SH-PEDOT)

To prepare the stretchable and conductive PEDOT/Au hybrid interconnects, the base PEDOT layer needs to have excellent stretchability and electrical performance. Poly(ethylene glycol) diacrylate (**PEGDA**) has been previously used as a polymeric additive in PEDOT:PSS to promote PEDOT aggregation and enhance its conductivity and stretchability (**PEG-PEDOT**) (**Fig. 2A (i) -2B (i)**) (*22*, *23*). However, the tendency for crystallization of linear PEGDA chains limits the stretchability of films typically to <50% strain. To overcome this issue, we previously reported a molecular design to suppress PEG crystallization by threading it with rotaxanes, however, this approach requires multi-step synthesis processes (*23*). In addition, PEG-PPG-PEG diacrylate (P123DA) blended into PEDOT can improve stretchability, but the conductivity is substantially lower (*24*).

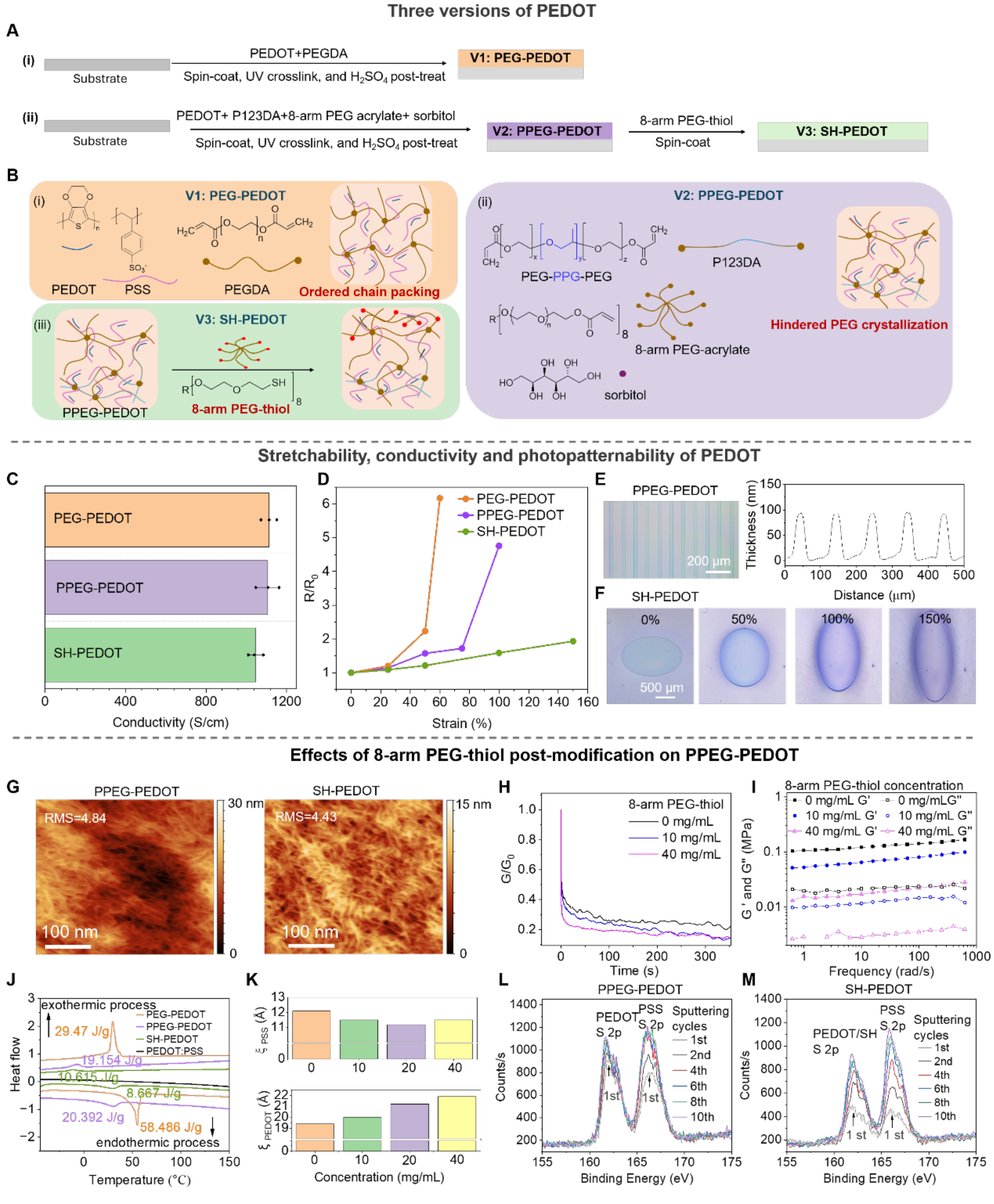

**Figure 2. Composition and morphological characterization of the highly stretchable and conductive SH-PEDOT films** (**A**) Fabrication processes of three types of PEDOT films. (**B**) Chemical compositions and the schematic morphology of the PEDOT films. (**i**) PEG-PEDOT: PEGDA induces PEDOT aggregation and provides a crosslinked matrix with modest stretchability due to its crystallization. (**ii**) **PPEG-PEDOT**: P123DA coupled with PPG blocks can reduce PEG crystallization and improve film stretchability, but reduces conductivity. Therefore, the 8-arm PEG-acrylate and sorbitol are added to promote PEDOT aggregation while maintaining high stretchability. (**iii**) **SH-PEDOT**: post-modification of **PPEG-PEDOT** with 8-arm PEG-thiol further reduces PEG crystallinity, disturbs PSS packing while promoting further PEDOT aggregation; the thiol molecules are more concentrated in the upper region of the film. (**C**) Conductivity of PEDOT films measured by four-probe conductivity measurements. The error bars show the standard error for three samples. (**D**) **SH-PEDOT** exhibits a higher stretchability than **PPEG-PEDOT** and **PEG-PEDOT**. These films (~ 100 nm thickness) were spin-coated on poly(dimethylsiloxane) (PDMS) substrates. R: resistance. $R_0$: initial resistance. (**E**) Direct photopatterned **PPEG-PEDOT** and its thickness profile measured by a profilometer. (**F**) Optical images of **SH-PEDOT** showing it can be stretched up to 150% strain without visible cracks. (**G**) AFM topographical height images of **PPEG-PEDOT** and **SH-PEDOT** films, showing that the fibrous PEDOT structures are preserved after the 8-arm PEG-thiol post-treatment. (**H**) Stress relaxation measurements of **PPEG-PEDOT** and **SH-PEDOT** with different amounts of 8-arm PEG-thiol, showing faster stress relaxation with the post-treatment. (**I**) Storage moduli (G') and loss modulus (G'') of **PPEG-PEDOT** and **SH-PEDOT** with different amounts of 8-arm PEG-thiol. (**J**) DSC heating and cooling curves of **PEG-PEDOT**, **PPEG-PEDOT**, **SH-PEDOT**, and **PEDOT:PSS**. (**K**) Coherence length ($\xi$) of PSS (upper panel) and PEDOT (lower panel) domains with increasing concentration of 8-arm PEG-thiol used for post-modification, extracted and calculated from GIWAXS Full Width at Half Maximum. (**L**) XPS sulfur depth profiles of **PPEG-PEDOT** and (**M**) **SH-PEDOT** films.

Here, to reduce the tendency for crystallization, our base layer stretchable PEDOT (**PPEG-PEDOT**) was prepared by incorporating an 8-arm Poly(ethylene glycol) acrylate (**8-arm PEG-acrylate**) into a PEDOT:PSS solution. Also, we employed P123DA to ensure stretchability and added sorbitol to enhance conductivity through promoting solution aggregation of PEDOT (**Fig. 2A (ii) -2B (ii)**). The PPG segments in P123DA are hydrophobic and sterically hindered, which can effectively suppress crystallization and improve mechanical stretchability (**Fig. 2B (ii))**. However, a high fraction of P123DA deteriorates conductivity. Therefore, the 8-arm PEG-acrylate and sorbitol are introduced to help promote aggregation of PEDOT:PSS in the solution state and enhance the resulting film's conductivity. Sorbitol further acts as a secondary dopant as well as a plasticizer for PSS to boost both conductivity and stretchability (*25*). After sulfuric acid post-treatment and water washing, sorbitol was effectively removed from the PEDOT film. The resulting **PPEG-PEDOT** thin film showed a high conductivity of 1164 S/cm and can be stretched to 100% strain without cracks (**Fig. 2C-2D**). Last, we showed that it can also be directly photopatterned (**Fig.2E**).

Next, we post-treated the crosslinked **PPEG-PEDOT** film with an 8-arm PEG-thiol to prepare **SH-PEDOT** (**Fig. 2A (ii) -2B (iii)**). This treatment introduces thiol functional groups into the PEDOT network, which may potentially strengthen PEDOT-gold adhesion and improve gold stretchability. Previously, PEG post-treatment of PEDOT film has been shown to further enhance PEDOT aggregation (*26*). Here, we hypothesized that the PEG-based 8-arm PEG-thiol may also readily be incorporated into **PPEG-PEDOT** through post-treatment due to their chemical structure similarities, while simultaneously providing abundant thiol sites for bonding with gold. Importantly, 8-arm PEG-thiol is commercially available, which makes our stretchable electrodes readily accessible.

Our obtained results indicated that **SH-PEDOT** exhibited a comparable conductivity to **PPEG-PEDOT**, while showing a crack-onset strain exceeding 150% and more stable resistance response under applied strain (**Fig. 2C-2D, Fig. 2F)**. To understand the origin of this excellent performance, we first used atomic force microscope (AFM) to image these films, in which a fibrous morphology of PEDOT was observed, indicating PEDOT aggregations remained after the addition of 8-arm PEG-thiol. (**Fig. 2G**). Such fibrous PEDOT morphology, sometimes called nanoconfined PEDOT morphology, has previously been linked to high conductivity due to the presence of ordered and interconnected PEDOT structures (*27*, *28*). Next, we investigated the roles of the 8-arm PEG-thiol on the mechanical properties and packing of **PPEG-PEDOT** films. Our obtained rheological data showed that, with an increasing amount of 8-arm PEG-thiol, **PPEG-PEDOT** films exhibited faster stress relaxation and lower moduli (**Fig. 2H-2I**), indicating increased chain mobility and softening of the film by 8-arm PEG-thiol. Differential scanning calorimetry (DSC) showed that, as compared to **PEG-PEDOT** and **PPEG-PEDOT**, the **SH-PEDOT** exhibited substantially reduced PEG crystallinity as evidenced from the lower heat of melting and heat of recrystallization (**Fig. 2J**). In addition, grazing-incidence wide-angle x-ray scattering (GIWAXS) characterizations of the thin films revealed that with more 8-arm PEG-thiol, **PSS** exhibited a shorter coherence length, indicating a reduced PSS crystalline domain size, while PEDOT showed a longer coherence length, suggesting enhanced long-range crystalline ordering (**Fig. 2K**). Using Fourier transform infrared spectroscopy (FTIR), we further showed that the incorporation of 8-arm PEG-thiol shifted the **PSSH**'s S=O stretching vibrations to higher wavenumbers, which may be caused by either H-bonding breaking within **PSSH** or more disorder in packing (consistent with reduced coherence length from GIWAXS). All these possibilities will consequently reduce the modulus of the **PSS** matrix.

Next, X-ray photoelectron spectroscopy (XPS) measurement was used to determine the vertical distribution of 8-arm PEG-thiol within the film. PEDOT, PSS, and 8-arm PEG-thiol each have a sulfur signal at around 161 eV, 166 eV, and 161eV, respectively. **Fig. 2L-2M** show that in **SH-PEDOT** and **PPEG-PEDOT**, sulfur signals from PEDOT and PSS are both detectable on film surfaces (i.e., at the beginning of the sputtering cycle). This suggests that PEDOT:PSS is at least not fully covered by a layer of 8-arm PEG-thiol, which would have shown a single sulfur signal at around 161 eV-162 eV. However, unlike **PPEG-PEDOT**, which shows a small change in sulfur signals across sputtering cycles suggesting even distribution of PEDOT throughout the film, for **SH-PEDOT,** both PEDOT and PSS sulfur signals increase with sputtering cycles. This indicates that PEDOT content increases with film depth and the 8-arm PEG-thiol molecules reside more in the upper region of the film.

Collectively, the above findings confirmed that 8-arm PEG-thiol addition to **PPEG-PEDOT** softens the film due to reduced crystallinity of PEG and disturbed packing between PSSH. These factors contributed to the improved stretchability of **PPEG-PEDOT**. Since PEDOT packing improved as evidenced from a larger crystalline domain coherence length, this indicated the electrical conduction pathways have not been disturbed and explained our observed excellent electrical performance. Furthermore, since 8-arm PEG thiol molecules resided more in the upper region, this provided the thiol groups for gold bonding. Therefore, **SH-PEDOT** provides an excellent stretchable and conductive layer for subsequent gold deposition.

## Highly stretchable and conductive SH-PEDOT/Au interconnect lines with narrow linewidths

To enable highly conductive interconnect lines, we further deposited a thin Au layer (20-30 nm) onto the thiol-rich **PPEG-PEDOT** to prepare the stretchable, conductive electrode with metal-level conductivity (**SH-PEDOT/Au**) (**Fig. 3A**). The Au layer greatly increased the overall conductivity of the PEDOT film from ~1,000 S cm$^{-1}$ to >42,700 S cm$^{-1}$ (**Fig. 3B**). The resulting **SH-PEDOT/Au** film was stretchable to 170% strain, without any visible cracks under an optical microscope. AFM images revealed that only sub-micron cracks were formed in **SH-PEDOT/Au** film (**Fig. 3C**), which provides the opportunity for micro-size interconnect lines. In contrast, the **PPEG-PEDOT/Au** film (which lacks the 8-arm PEG-thiol post-treatment) formed large cracks at 100% strain (**Fig. 3C**). The resulting conductivity of **SH-PEDOT/Au** was maintained >14760 S/cm at 100% strain, while the **PPEG-PEDOT/Au** film dropped to 3830 S/cm. These observations indicated that the interfacial bonding between Au and thiols in the **SH-PEDOT** layer effectively suppressed crack propagation and prevented the formation of large cracks. Such a strategy reinforced the previous observation that interfacial cohesion can improve ductility of metal films (*29*).

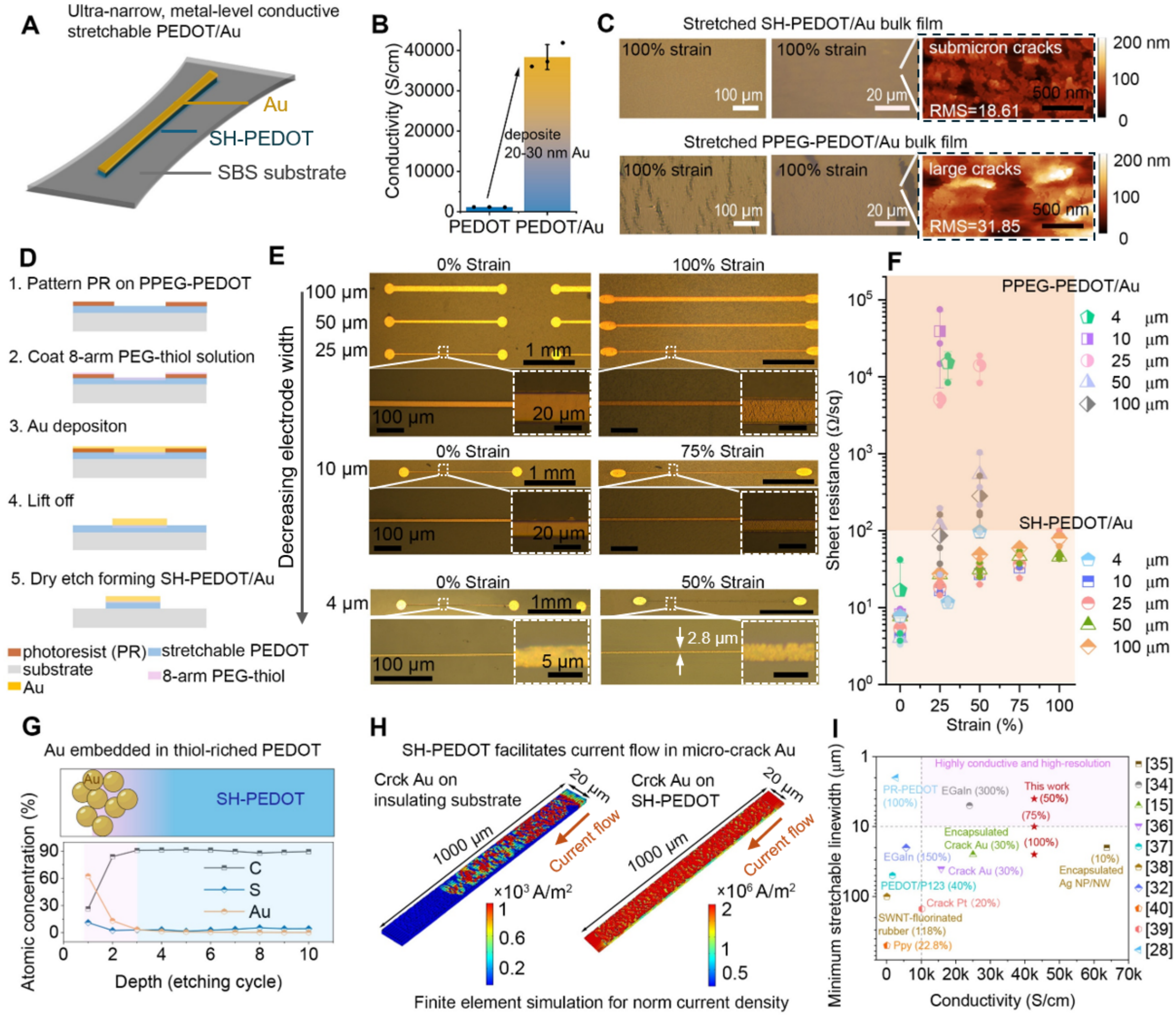

**Figure 3. Highly conductive SH-PEDOT/Au interconnects with excellent stretchability at narrow line widths.** (**A**) The schematic illustration of a patterned stretchable **SH-PEDOT/Au**

stack structure. (**B**) Conductivity of **PPEG-PEDOT** significantly increases after deposition of a thin Au layer (20 nm–30 nm). (**C**) Optical microscope images of **SH-PEDOT** and **PPEG-PEDOT** films under 100% strain. The magnified views show corresponding AFM images. In the optical microscope image, **PPEG-PEDOT** exhibits large cracks, whereas **SH-PEDOT** shows no visible cracks. AFM images reveal submicron cracks in **SH-PEDOT** and large cracks in **PPEG-PEDOT** under 100% strain. (**D**) Patterning processes of **SH-PEDOT/Au** lines using standard photolithography. (**E**) Optical microscopic images of **SH-PEDOT/Au** interconnect with widths from 100 µm to 4 µm under 100% to 50% strains. Magnified views are shown below and in the white dashed box. Scale bars from left to right columns of images use the same scales. Their sheet resistance change as a function of strain is shown in (**F**). (**G**) XPS depth profile of **SH-PEDOT/Au**, showing the coexistence of C, Au, and S elements in the first three etching cycles. (**H**) COMSOL finite element simulation of current density in cracked Au morphology from 20 µm at 100% strain on an insulating layer (left) and on **SH-PEDOT** (right, assuming 1000 S/cm). The **SH-PEDOT/Au** system shows a uniform and much higher current density throughout the film, while the conduction pathways are mostly disrupted when supported on an insulating substrate. (**I**) Comparison of the minimum reported widths of stretchable electrodes and their conductivity with our work. The numbers in bracket indicate the stretchability achieved at the corresponding widths. The purple shading indicates materials that achieve metal-like conductivity and high-resolution (<10 µm). The error bars in (**B**) and (**F**) show the standard error for three samples.

While cracked-Au film has been widely used for stretchable electronics (*30*, *31*), fabricating high-resolution stretchable sub-100 micron narrow interconnects remains challenging. To maintain conduction pathways, the crack dimension should be substantially smaller than the interconnect linewidth, such that there is no complete current path interruption. In previous reports, only liquid metal lines showed stable resistances under >100% strain at widths down to 10 µm (*32*, *33*). Their susceptibility to both oxidation and moisture degradation remains a concern. For instance, we observed that 10 µm EGaIn lines showed >60% increase in resistance after 6 days in a humid chamber, degrading much faster than wider lines (100 µm), likely due to their higher surface-area-to-volume ratio and greater moisture exposure.

We show that our SH-PEDOT/Au films can be readily patterned into lines with linewidths as small as 1.5 µm using standard photolithography and etching processes (**Fig. 3D**) or patterned using a shadow mask. Importantly, the patterned 25-100 µm linewidth can be stretched up to 100%-150% strain, the 10 µm linewidth can be stretched up to 75%, while the 4 µm linewidth can be stretched up to 50% strain with minimal cracks under optical microscope and able to maintain sheet resistances lower than 100 Ω/sq (**Fig. 3E-3F**). In contrast, **PPEG-PEDOT/Au** interconnects formed large cracks and sheet resistances increased >1 kΩ/sq at a low strain of 25% for 25-µm lines (**Fig. 3F**) and increased substantially more for 100-µm and wider lines under high strains (**Fig. 3F**).

Next, we use XPS depth profile to characterize the distribution of Au in the **SH-PEDOT**/Au film (**Fig. 3G**). In the first three etching cycles, signals from carbon (C), sulfur (S), and gold (Au) were simultaneously detected. As etching progressed, the C signal became increasingly dominant, while the S signal remained weak and the Au signal eventually diminished. This suggests that there is an interfacial region consisting of a mixture of both Au and **SH-PEDOT**, while some of the Au formed Au-thiol bonds in this region. Such an interfacial structure provides both physical interlocking and chemical bonding between Au and SH-PEDOT, thereby enhancing interfacial adhesion. Although in PPEG-PEDOT/Au XPS depth profile, C and Au were also detected in the first three etching cycles, suggesting a similar interlocking region, its S signal was substantially weaker, indicating the absence of thiol-Au bonds at the interface. Hence, the combination of interfacial interlocking and Au–thiol bonding enhances adhesion between Au and SH-PEDOT, constraining crack opening under strain and resulting in smaller Au crack sizes. Compared to the previous stretchable electrodes based on cracked Au, our SH-PEDOT/Au shows reduced crack sizes, which is essential to produce narrow linewidth electrodes.

In our design, another important advancement is the use of a stretchable and highly conductive PEDOT layer to help bridge the conductive pathways in the micro-cracked Au regions. We used COMSOL finite-element simulation to understand current flow distribution across related films (**Fig. 3H**). In the simulation, we extracted the crack morphologies from a 20 µm-wide Au electrode under 100% strain and modeled its current flow on either an insulating substrate or 8-arm PEG-thiol-treated **SH-PEDOT.** We observed that current flow along the conductive pathway in the film was greatly disrupted when it was placed on an insulating substrate. In contrast, the current flow with the **SH-PEDOT** substrate remained much more uniform throughout the length of the film. This indicates that having a conductive PEDOT layer helps in maintaining current flow, which led to a much more stable conductivity as observed in narrow electrodes under large strains. Additionally, both our 100-µm and 20-µm wide electrodes showed stable performance when subjected to 2000 cycles at 50% strain. Furthermore, our electrodes showed negligible impedance changes after 12 days of accelerated aging in PBS (equivalent to 48 days at room temperature) and nearly unchanged conductance after buffer soaking and subsequent storage in air for 15 months. In contrast, narrow EGaIn interconnects rapidly degraded after 18 hours in PBS solution, even when encapsulated with a 1-µm SBS layer.

Furthermore, we emphasize that our stretchable conductor design is applicable to other metals, including Ag, Pt, and Pd, as they can all bond to thiol groups. For instance, 50 µm-wide **SH-PEDOT/Ag** interconnects showed negligible crack formation and resistance change up to 75% strain, whereas untreated samples resulted in large cracks at 25% strain. Thiol groups can be potentially replaced with other functional groups to enable additional metal thin films to be incorporated into stretchable electronics. We further observed that upon adding an additional **PPEG-PEDOT** layer onto **SH-PEDOT/Au** (thus forming a sandwiched structure) can enhance electrical stability under strain. Our prepared tri-layered films showed negligible changes in sheet resistance even at 100% strain. Also, these patterned sandwich structures can be readily fabricated using standard lithography processes. Hence, our **SH-PEDOT/Au** interconnects stand out among currently reported stretchable conductors, demonstrating high metal-level conductivity while maintaining stretchability with widths as small as 4 µm (**Fig. 3I**) (*15*, *28*, *32*, *34–40*).

### Stretchable transparent interconnects for high-density arrays

As a key advancement, we emphasize that our highly stretchable and conductive **SH-PEDOT/Au** can be used as transparent interconnects for device arrays. For example, stretchable LED displays have been actively pursued for displays, wearable electronics, smart textiles, and XR/AR devices. The interconnects must possess high conductivity to support stable, high current densities over long distances. At the same time, they must have narrow linewidths to enable high pixel density (PPI) and optical transparency. Recently, companies have introduced stretchable micro-LED displays with ~200 PPI and up to 30% stretchability using metal serpentine structures as stretchable interconnects (*41*). As pixel density further increases, intrinsically stretchable transparent interconnects are needed. However, none of the previously reported materials can meet the stringent requirements.

Transparent conductors can be made from oxide films, metal nanowire meshes, and patterned grids. In this work, we patterned **SH-PEDOT/Au** into grid structures with a width of 25 µm (W) and varying pitch sizes (P) (**Fig. 4A**), which results in fill factors (FF = W/P) from~ 4% to 13%. These samples showed high transparency across the visible range. At a fill factor of 13%, it showed 92% optical transparency at 550 nm (**Fig. 4B**) and a sheet resistance of 17.8 Ω/sq and 100% stretchability (**Fig. 4C**). Among existing transparent conductors, only ITO and metal grids can achieve both high optical transmittance (>90%) and low sheet resistance (<30 Ω/sq). Here, we note that our SH-PEDOT/Au grid is able to rival these two materials, while offering a much higher stretchability (**Fig. 4D**). Although EGaIn grids exhibit extremely low sheet resistance, their optical transmittance is often below 90%. Furthermore, they usually showed a more pronounced optical diffusion and reflection due to their rough surface morphology and large thickness. It’s worth noting that increasing the gold thickness can further reduce the sheet resistance of our SH-PEDOT/Au grid.

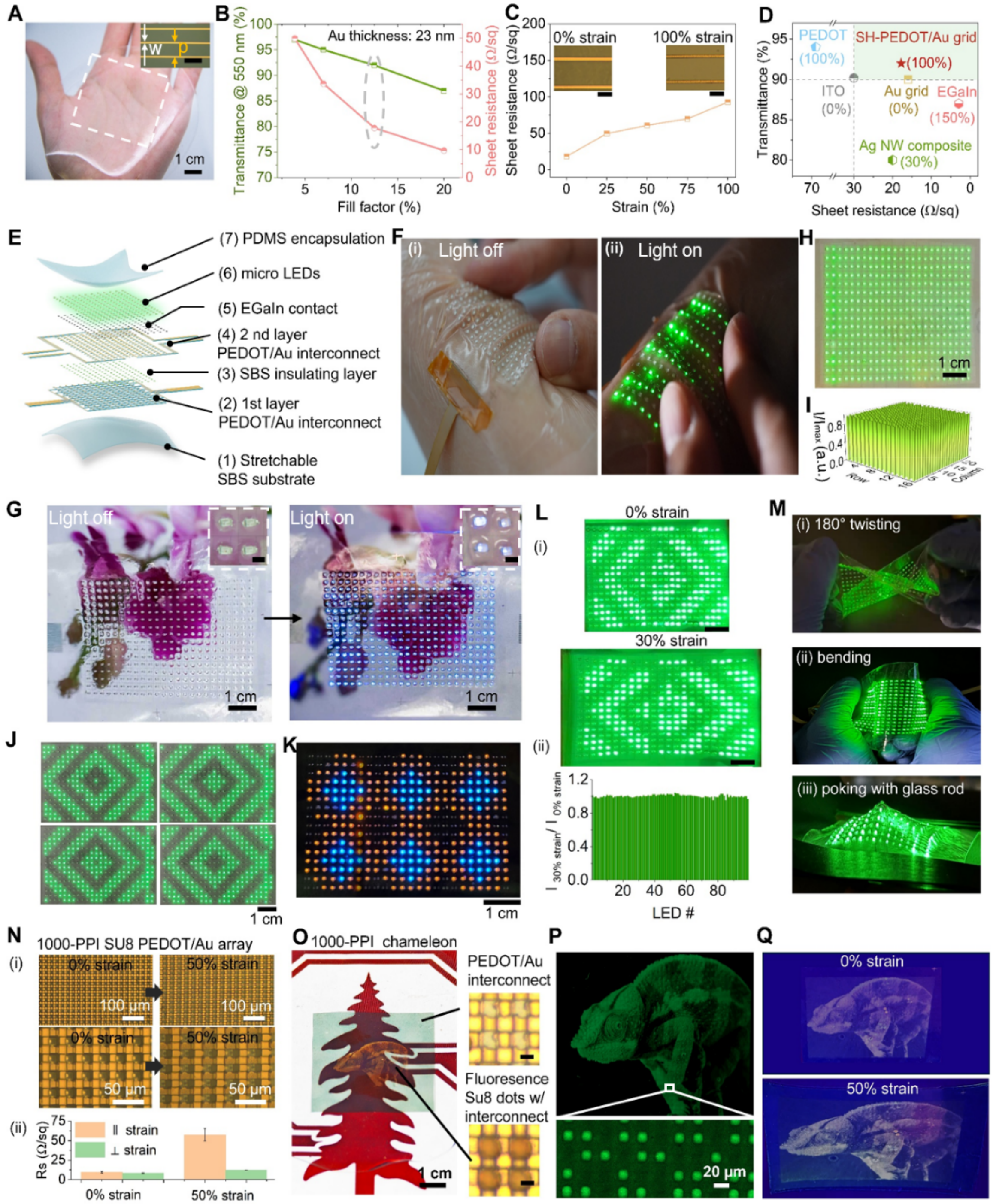

**Figure 4. Stretchable micro-LED display enabled by high-density transparent SH-PEDOT/Au interconnects**. (**A**) Photograph and optical microscope image (inset) of the **SH-PEDOT/Au** grid with fill factor of ~13%. W denotes the interconnect width and P the interconnect pitch. They are 25 µm and 200 µm, respectively. Scale bar: 200 µm. The fill factor (FF) is defined as FF = W/P. (**B**) Transmittance at 550 nm and sheet resistance of **SH-PEDOT/Au** grids with varying fill factors. (**C**) Sheet resistance changes of **SH-PEDOT/Au** grids under strain with a fill factor of 13 %. The inset image shows the optical microscopic image of the grid at both 0% and 100% strain. Scale bar: 50 µm. **(D)** Comparison of sheet resistance and optical transmittance of different types of transparent conductors. The number alongside each conductor in bracket indicates its stretchability. (**E**) Schematic illustration of the seven-layer micro-LED array. (**F**) Photograph of the conformable device adhered to skin in the off- (**i**) and on-states (**ii**). (**G**) Photograph of the micro-LED array in the off- and on- states placed over flowers, in which the underlying flower is clearly visible. Note the flower is defocused in the image, but is focused on the array. The inset shows a magnified view of the device, where the interconnects are barely noticeable. Scale bar in inset: 500 µm. (**H**) Photograph of an array with all micro-LEDs illuminated, demonstrating uniform light intensity. (**I**) Normalized luminescence ($I/I_{max}$) of all 384 micro-LEDs in the array, showing uniform brightness distribution. (**J**) Snapshot images extracted from a video showing the array displaying dynamic images (**Movie S3**). (K) Photo of an array with multi-color micro-LEDs. (**L**) (i) The micro-LED array maintains its original brightness under 30% tensile strain. Scale bar: 1 cm. (ii) Relative micro-LED emission intensity under 30% strain, showing minimal degradation. (**M**) The array maintains stable operation under (**i**) 180-degree twisting, (**ii**) bending, and (**iii**) poking. (**N**) (i) Optical microscopic images of 1000-PPI SU8 islands patterned with SH-PEDOT/Au interconnect before and after 50% strain. Interconnect width: ~10 µm, Island width: ~12 µm. (ii) Sheet resistance changes of the interconnect parallel and perpendicular to the stretching direction before and after 50% strain. (**O**) A 1000-PPI chameleon image, which was patterned using fluorene polymer embedded within SU-8 islands. The background picture is clearly visible, demonstrating high transparency of the device. The magnified view shows the optical microscopic image of the background region containing only the **SH-PEDOT/Au** interconnect lines, and the chameleon region where fluorescent/SU8 islands overlay the interconnects; both regions exhibit high transparency. Island width: ~15 µm, interconnect width: ~10 µm. Scale bar: 10 µm. (**P**) Fluorescence microscopic image of the 1000-PPI chameleon image, with magnified views showing fluorescent dots over the interconnects. (**Q**) The image remains intact under 50% strain.

To show the capability of our **SH-PEDOT/Au** interconnects, we next fabricated a passive transparent 24 × 16 micro-LED matrix with a customized printed circuit board (PCB) and a microcontroller (**Fig. 4E**). The fabrication process of our interconnect is highly scalable and reliable, where measured trace resistances across a four-inch wafer closely matched the calculated resistance values. Our device, excluding the micro-LEDs, is as thin as ~5 µm and shows excellent skin conformability (**Fig. 4F, Movie S1**). Besides, the array is transparent except the micro-LEDs regions, using interconnects with 30 µm linewidths (**Fig. 4G, Movie S2).**

The transparent micro-LED array maintained consistent luminance over a large area (20 cm, the longest line in the array) (**Fig. 4H-Fig.4I**). Additionally, the current switches on and off even at >1 kHz, allowing the passive matrix to display dynamic and colorful information through real-time programming (**Fig. 4J-4K, Movie S3-S5**). The array can sustain stretch (30% strain) while maintaining stable LED lighting performance (**Fig. 4L**). It remained functional even under various mechanical deformations, including poking, twisting, bending, and crumpling (**Fig. 4M, Movie S6**).

Our developed high-density interconnects will enable next-generation, transparent high-PPI stretchable micro-LED displays. In the absence of advanced micron-sized LEDs, we fabricated a 1000-PPI model array in which two layers of ~10-µm-wide SH-PEDOT/Au interconnects were overlaid with densely packed 12-µm SU-8 rigid microdots that served as surrogate micro-LED pixels (**Fig. 4N**). Despite the extremely high pattern density, the 1000-PPI interconnects maintained sheet resistance <70 Ω/sq under 50% strain. It is worth noting that the integration of rigid islands presents a significant challenge for stretchable interconnects, as the rigid regions deform minimally and force the interconnects to absorb most of the applied strain. Nevertheless, our interconnects remained functional and exhibited only modest rigid-island-induced increases in sheet resistance compared with those without rigid islands under such high density. This robustness likely arises from the presence of PEDOT, which maintains charge transport across strain-induced cracks in the Au layer and preserves electrical performance under severe local strain. This property makes SH-PEDOT/Au an attractive platform for high-PPI stretchable micro-LED displays, a highly sought-after technology that remains challenging to achieve. To further demonstrate this capability, we patterned fluorescent/SU-8 composite pixel dots together with SH-PEDOT/Au interconnects to create a 1000-PPI chameleon image (**Fig. 4O-4P**). The fluorescence image and the interconnect network show high optical transparency (**Fig. 4O**) and can also withstand 50% tensile strain without any noticeable cracking under optical microscope (**Fig. 4Q**).

## Discussion

We note that as rigid pixel density increases, the stretchability requirements for the interconnects increase correspondingly. This requirement becomes even more stringent with larger rigid island sizes, which underscores the need for a highly stretchable interconnect. At the same time, for ultra-high pixel densities, the interconnects must also have narrower linewidths to accommodate the limited footprint. Structural engineering using wavy or serpentine structures is unable to provide sufficient space to meet both the footprint and stretchability requirements. However, our interconnect strategy provides avenues for high-PPI, stretchable, and transparent micro-LED displays.

We further emphasize that our interconnects combine the complementary advantages of both Au and PEDOT, thereby outperforming either material alone. For instance, in addition to the high conductivity imparted by Au, the presence of PEDOT provides low electrochemical impedance. As a result, for bioelectronic applications, SH-PEDOT/Au interconnects exhibit lower impedance, higher volumetric charge-storage capacity, and higher charge-injection capacity than either bare Au or PEDOT alone, highlighting their broader functionality and application potential compared with the individual constituent materials.

In conclusion, through rational material and interfacial engineering, we developed a stretchable conductor that can be patterned at high resolution using standard photolithographic processes. The obtained conductive lines exhibit metal-level conductivity (>40,000 S/cm), maintain stretchability at linewidths as small as 4 μm, and remain stable for over 15 months in air, humid environments, and PBS. This combination of conductivity, stretchability, patternability, and environmental stability establishes a materials platform for high-density stretchable electronics that has not previously been achieved. As demonstrations, we show applications in transparent conductors and high-PPI micro-LED displays, performance regimes that are unattainable with existing stretchable conductor systems. We anticipate that this technology will expand the capabilities of stretchable electronics and enable new opportunities in bioelectronics, displays, wearable systems, and soft robotic devices.

**Acknowledgments:** We thank Asahi Kasei Co. for providing the SEBS polymer.

**Funding:** This work was supported by Stanford wearable electronics initiative seed funding, eWEAR Chen Ideation and Prototyping lab.

**Author contributions:**

K.C. and Z.B. conceived the concept and designed the project. K.C., E.T.Z. designed the experiments, carried out the device fabrication, and collected data. W.W. and Y.Z provided critical ideas. K.C. carried out PEDOT/Au interfacial engineering design and fabrication. E.T.Z. and K.C. carried out stretchable micro-LED circuit design and fabrication. K.C. and W.W. carried out PEDOT design and characterization. Y.Z. and K.C. carried out stretchable micro-LED fabrication. K.P. carried out P123DA polymer synthesis. L.F. carried out AFM measurements. Y.S. carried out rheology measurements. Tiany.C. and E.Z. carried out depth XPS measurements. R.W. carried out GIWAX measurements. Tianh.C. and K.C. carried out stability experiments. H.L., S.W., J.Z., W.W., helped with photography and provided invaluable experiment suggestions. Yat.Y. helped with interconnect fabrication. Yuj. Y helped with stretchable micro-LED and EGaIn interconnect fabrication. R.R.Z. helps with stretchable micro-LED fabrication and literature sorting. K.C. carried out COMSOL modeling. K.C., E.T.Z., W.W., J. T., and Z.B. wrote the manuscript. All authors reviewed and commented on the manuscript.

**Competing interests:** Stanford University has filed a patent application related to this technology. The patent application number will be added in the future.

**Data, code, and materials availability:** All data are available in the main text or the supplementary materials.